\documentstyle[aps]{revtex}

\begin{document}

\input{epsf}

%\draft

\title{THE WEAKLY DISORDERED 2D ELECTRON GAS IN A MAGNETIC FIELD}
%      =========================================================
\author{N. Brali\'c\cite{NB}}
\address{Departamento de F\'\i sica, Pontificia Universidad
Cat\'olica de Chile \\
Casilla 306, Santiago 22, Chile}
\author{R. M. Cavalcanti\cite{RMC}}
\address{Departamento de F\'\i sica, Pontif\'\i cia Universidade
Cat\'olica do Rio de Janeiro \\
CP 38071, CEP 22452-970, Rio de Janeiro, RJ, Brasil}
\author{C. A. A. de Carvalho\cite{CAAC} and P. Donatis\cite{PD}}
\address{Instituto de F\'\i sica, Universidade Federal do
Rio de Janeiro \\
CP 68528, CEP 21945-970, Rio de Janeiro, RJ, Brasil}
\date{April 10, 1997}
\maketitle

\begin{abstract}
%      --------

We study a relativistic two-dimensional free electron gas in the
presence of a uniform external magnetic field and a random static
scalar potential.  We compute the expectation values of the charge
density and the conductivity tensor, averaged over the random
potential in first order perturbation theory.  We discuss the
effect of a zero vs. finite range correlation length of the random
potential.

\end{abstract}

\section{Introduction}
%        ------------

The free two-dimensional electron gas (2DEG) (without electron-electron
interactions) in the presence of a uniform external magnetic field has
become a system of great interest.  From a field theoretical point of
view, it is the starting point for a non-perturbative treatment of
QED$_{2+1}$\cite{QED3,Chodos}, while in condensed matter it is the basic
framework for the discussion of the (integer-) Quantized Hall Effect
(QHE)\cite{IQHE}.  One of the salient features of the system is its
discrete density of states consisting of equally degenerate Landau
levels.  They give rise to some of the key features of the 2DEG, but
are also responsible for much of the technical difficulties.

Both in field theory and in the applications to the QHE, it is widely
believed that the spreading of the energy levels into Landau bands of
finite width may have profound physical consequences, through the
coexistence of localized and delocalized states.  In this context, it
is expected that the effect of disorder due to impurities or
fluctuations of the magnetic field background might play an essential
role.  At the same time, however, it is necessary to preserve some
universal features such as the precise unit of quantization of the Hall
conductivity.

In this paper, we analyze the effects of disorder introduced by a random
``scalar'' potential (as opposed to disorder due to random fluctuations
in the background magnetic field).  We compute the expectation values
of the charge density $\langle j_0 \rangle$ and of the conductivity
tensor $\langle \sigma_{ij} \rangle$ averaged over the random potential.
Although one can put forward general arguments that suggest that the
effect of disorder is essentially non-perturbative, we perform our
calculation in a first order perturbation expansion.  There are several
reasons to do so.  First, to the best of our knowledge, a direct field
theoretic calculation from first principles has not been done.  Second,
whatever is or is not obtained in a perturbative calculation may shed
some light as to what must be sought in subsequent non-perturbative
treatments.  Third, we find that even a perturbative calculation gives
non-trivial results when one studies the effect of different correlation
lenghts of the random field.

The paper is organized as follows.  In section \ref{NoDisorder}, we
summarize the results for the charge density and the conductivity
of the free 2DEG in the absence of disorder.  Those are well known
results but we take the opportunity to fix our notation and to discuss
some important aspects of their physical interpretation which, in
our opinion, are somewhat confusing in the literature.  Next, in
section \ref{ZeroCorrelation}, we present the perturbative calculation
of the effects due to a scalar random potential with a Gaussian
distribution of {\it zero\/} correlation length.  Somewhat surprisingly,
we find that, although the relation between the conductivity and the
charge density remains unchanged, their quantization units in terms
of the magnetic field is modified.  Then, in section
\ref{FiniteCorrelation}, we discuss the modification of the preceeding
calculation in the case of a finite correlation length, showing that,
at least in this perturbative approach, the disorder has {\it no\/}
observable effects for any finite value of its correlation length.
Finally, in section \ref{Conclucoes}, we summarize our results and
discuss their implications.

\section{Charge density and conductivity in the absence of disorder}
%        ----------------------------------------------------------
\label{NoDisorder}

We shall study an electronic gas in $(2+1)$ dimensions in the presence
of a uniform magnetic field and a random (scalar) potential.  Its
Lagrangian is given by
\begin{equation}
{\cal L} =
\overline\psi\,
(
  i\partial\!\!\!\slash+eA\!\!\!\slash-m+\mu\gamma^0
)\,\psi  \; .
\label{Lagrangian}
\end{equation}
Here the field $A^{\mu}$ stands for $A^{\mu}=(V,{\bf A})$, where the
vector potential ${\bf A}$ accounts for the uniform magnetic field
$B\,(=\partial_1A^2-\partial_2A^1)$, and $V$ is a static random
potential, describing quenched disorder.  $\mu$ is the chemical
potential, and the $\gamma$-matrices are Pauli matrices:
$\gamma_0=\sigma_3$, $\gamma_{1,2}=i\sigma_{1,2}$, satisfying
$\{\gamma_{\mu},\gamma_{\nu}\}=2\eta_{\mu\nu}{\bf 1}$, with
$\eta_{\mu\nu}=diag(1,-1,-1)$.

The charge density $j_0(x)$ is given by
\begin{equation}
j_0(x) = ie \, {\rm Tr} \, [\gamma_0\,S(x,x)] \; ,
\label{j_0}
\end{equation}
where $S(x,y)$ is the Feynman propagator of the theory, satisfying
\begin{equation}
(i\partial\!\!\!\slash_x+eA\!\!\!\slash(x)-m+\mu\gamma^0)\,S(x,y)=
\delta^3(x-y)  \; .
\label{feynman}
\end{equation}

If a perturbing electric field ${\bf E}$ is turned on, a current $J_i$
will be induced which, in the linear response regime, is given by
\begin{equation}
J_i(x)=\int d^3y\,\Pi_{i\nu}(x,y)\,{\cal A}^{\nu}(y)  \; ,
\end{equation}
where $\Pi_{\mu\nu}(x,y) = -ie^2\,{\rm Tr}\,[\gamma_{\mu}\,S(x,y)\,
\gamma_{\nu}\,S(y,x)]$ is the polarization tensor and ${\cal A}^0(x) =
-{\bf E}\cdot{\bf x}$, ${\cal A}^i(x)=0$. Thus, the D.C. conductivity
tensor, defined as $\sigma_{ij}(x) = \lim_{E\to 0}\,\partial J^i(x)/
\partial E^j$, is given by
\begin{equation}
\sigma_{ij}(x) = \int d^3y \, \Pi_{i0}(x,y)\,y^j  \; .
\label{sigma_ij}
\end{equation}

The charge density and the transverse conductivity of the system
without disorder ($V=0$) are well known
\cite{Lykken,Zeitlin1,Zeitlin2,DKK}.  They are given by (here $m$
and $eB$ are assumed positive; $\hbar=c=1$)
\begin{eqnarray}
j_0(V=0)
&=&
-\frac{e^2B}{2\pi}\,
\left\{
  \frac{1}{2}-\theta(-\mu-m)+\sum_{n=1}^{\infty}
  \left[\theta(\mu-\epsilon_n)-\theta(-\mu-\epsilon_n)\right]
\right\} \; ,
\label{j_0(V=0)}
\\
\sigma_{21}(V=0)
&=&
\frac{e^2}{2\pi}\,
  \left\{\frac{1}{2}-\theta(-\mu-m)+\sum_{n=1}^{\infty}
  \left[\theta(\mu-\epsilon_n)-\theta(-\mu-\epsilon_n)\right]
\right\} \; ,
\label{sigma_{21}(V=0)}
\end{eqnarray}
where $\epsilon_n\equiv\sqrt{m^2+2neB}$ are the relativistic Landau
levels.  Aside from an asymmetry, which is characteristic of the
relativistic theory, Eq.~(\ref{sigma_{21}(V=0)}) exhibits an integer
quantization of the transverse conductivity in units of $e^2/h$.
Indeed, restoring Planck's constant we have, for $\mu > 0$,
\begin{equation}
\sigma_{21}(V=0) = \frac{e^2}{2h} + n \,\frac{e^2}{h}
  \quad (n=0,1,2,\ldots)  \; .
\end{equation}
Furthermore, if we identify the chemical potential $\mu$ with the
Fermi energy $E_F$, the conductivity jumps from one plateau to the
next each time $E_F$ crosses a Landau level.  As a function of $E_F$,
in the non-relativistic limit, the plateaux have a fixed width of
$\hbar \omega_c$ ($\omega_c=eB/mc$ being the cyclotron frequency).
These features are characteristic of the integer QHE, but are far
from being {\em the\/} integer QHE.  Indeed, as shown by
Eq.~(\ref{j_0(V=0)}), the charge density as a function of $\mu$ is
also quantized, with {\em integer\/} filling factor. What is
happening is that, due to the zero width of the Landau levels, by
varying the chemical potential one can only adjust the number of
Landau levels which are occupied; but they are always {\em fully\/}
occupied.  In this way, what Eqs.~(\ref{j_0(V=0)}-\ref{sigma_{21}(V=0)})
show is that the transverse conductivity and the electron
concentration $\rho$ ( here $\rho = -j_0/e$) satisfy the {\em
classical\/} relation
\begin{equation}
\sigma_{21} = \frac{e}{B} \rho \; ;
\label{classical}
\end{equation}
however, as just mentioned, varying $\mu$ only allows us to reach
those points which satisfy $\rho = n eB/h$, with $n$ an integer.
The situation is best described in graphical terms (see Figure~1).

\section{Uncorrelated disorder}
%        ---------------------
\label{ZeroCorrelation}

To investigate the effect of disorder on the system, we must average
physical observables over all possible (static) configurations of
$V({\bf x})$, with a suitable weight.  In this section we choose an
uncorrelated Gaussian probability distribution:
\begin{equation}
\label{P}
P[V]=\exp\left\{-\frac{1}{2g}\int d^2x\,V^2({\bf x})\right\}  \; ,
\end{equation}
for which
\begin{equation}
\label{VV}
\langle V({\bf x})\rangle = 0 \; ,
\qquad
\langle V({\bf x})V({\bf y})\rangle =
  g \,\delta^2({\bf x}-{\bf y})  \; .
\end{equation}

Since the charge density $j_0$ and the conductivity tensor
$\sigma_{ij}$, given in Eqs.~(\ref{j_0},\ref{sigma_ij}) are
highly non-local functionals of $V$, we shall perform the averaging
perturbatively.  This is done by expanding the propagator $S$ in a
power series in $V$ and using Eq.~(\ref{VV}).  In matrix notation
($\Gamma_1(x,x') \equiv e\gamma_0 V(x)\,\delta^3(x-x')$):
\begin{equation}
S = S_0 - S_0 \Gamma_1 S_0 + S_0 \Gamma_1 S_0 \Gamma_1 S_0 + \ldots
\label{Dyson}
\end{equation}
The Feynman diagrams up to first order in $g$ for $\langle j_0(x)
\rangle$ and $\langle\Pi_{\mu\nu}(x,x')\rangle$ are depicted in
Figure~2.  Here $S_0$ is the unperturbed Feynman propagator,
satisfying Eq.~(\ref{feynman}) with $V=0$.  In regularized form,
it can be written as
\begin{equation}
\label{S_0}
S_0(x,y) =
M(x,y)\int dp^0\,e^{-ip^0(x^0-y^0)}\,\Sigma(p^0,{\bf x}-{\bf y}) \; ,
\end{equation}
where $\Sigma \equiv \Sigma_0\gamma^0 + \Sigma_1\gamma^1 +
\Sigma_2\gamma^2 + \Sigma_3{\bf 1}$, with
($\xi \equiv eB\,({\bf x}-{\bf y})^2/2$)
\begin{mathletters}
\label{Sigma}
\begin{eqnarray}
\Sigma_0(p^0,{\bf x}-{\bf y})
&=&
\frac{eB}{8\pi^2}\,e^{-\xi/2}\,
\sum_{n=0}^{\infty}
\left[
  \frac{p^0+\mu+m}{(p^0+\mu)^2-\epsilon_{n+1}^2} +
  \frac{p^0+\mu-m}{(p^0+\mu)^2-\epsilon_n^2}
\right]
L_n^0(\xi)\,e^{-\alpha n} \; ,
\\
\Sigma_1(p^0,{\bf x}-{\bf y})
&=&
-\frac{ie^2B^2}{4\pi^2}\,(x^1-y^1)\,e^{-\xi/2}\,
\sum_{n=0}^{\infty}
\frac{L_n^1(\xi)\,e^{-\alpha n}}{(p^0+\mu)^2-\epsilon_{n+1}^2} \; ,
\\
\Sigma_2(p^0,{\bf x}-{\bf y})
&=&
-\frac{ie^2B^2}{4\pi^2}\,(x^2-y^2)\,e^{-\xi/2}\,
\sum_{n=0}^{\infty}
\frac{L_n^1(\xi)\,e^{-\alpha n}}{(p^0+\mu)^2-\epsilon_{n+1}^2} \; ,
\\
\Sigma_3(p^0,{\bf x}-{\bf y})
&=&
\frac{eB}{8\pi^2}\,e^{-\xi/2}\,
\sum_{n=0}^{\infty}
\left[
  \frac{p^0+\mu+m}{(p^0+\mu)^2-\epsilon_{n+1}^2} -
  \frac{p^0+\mu-m}{(p^0+\mu)^2-\epsilon_n^2}
\right]
L_n^0(\xi)\,e^{-\alpha n}  \; ;
\end{eqnarray}
\end{mathletters}
$M(x,y)$ is a gauge dependent factor, given by
\begin{equation}
M(x,y)=\exp\left\{ie\int_{y}^{x}A_{\mu}(z)\,dz^{\mu}\right\}  \; ,
\end{equation}
where the integral must be performed along a straight line
connecting $y$ to $x$, and $L_n^a(z)$ $(a=0,1)$ are Laguerre
polynomials~\cite{GR}. The integral over $p^0$ in Eq.~(\ref{S_0})
must be performed along the contour depicted in Figure~3
\cite{Chodos,IZ}.  (Without the regulator $e^{-\alpha n}$, the sums
in Eq.~(\ref{Sigma}) are logarithmically divergent when ${\bf x}
={\bf y}$.  Thus, to avoid ambiguities, the limit $\alpha\to 0^+$
must be taken only at the end of the calculation.)

Now, we turn to the computation of the first order perturbative
correction to the charge density due to disorder. It is given by
\begin{eqnarray}
\label{j1}
\langle j_0(x)\rangle^{(1)}&=&ie^3\int d^3y\,d^3z\,{\rm Tr}\,[\gamma_0\,
S_0(x,y)\,\gamma_0\,S_0(y,z)\,\gamma_0\,S_0(z,x)]\,
\langle V({\bf y})V({\bf z})\rangle \nonumber \\
&=&ie^3g\int d^3y\,d^3z\,M(x,y)\,M(y,z)\,M(z,x) \int dp^0\,dq^0\,dk^0\,
e^{-ip^0(x^0-y^0)-iq^0(y^0-z^0)-ik^0(z^0-x^0)} \nonumber \\
&\times&{\rm Tr}\,[\gamma_0\,\Sigma(p^0,{\bf x}-{\bf y})\,\gamma_0\,
\Sigma(q^0,{\bf y}-{\bf z})\,\gamma_0\,\Sigma(k^0,{\bf z}-{\bf x})]\,
\delta^2({\bf y}-{\bf z})  \; .
\end{eqnarray}
Integrating over ${\bf z}$, $y^0$, $z^0$, $q^0$ and $k^0$, this
simplifies to
\begin{equation}
\langle j_0(x)\rangle^{(1)}=i4\pi^2e^3g\int d^2y\int dp^0\,{\rm Tr}\,
[\gamma_0\,\Sigma(p^0,{\bf x}-{\bf y})\,\gamma_0\,
\Sigma(p^0,{\bf 0})\,\gamma_0\,\Sigma(p^0,{\bf y}-{\bf x})]  \; .
\end{equation}
Evaluating the trace and performing the integral over ${\bf y}$, one
finds ($\omega\equiv p^0+\mu$)
\begin{eqnarray}
\label{j1a}
\langle j_0\rangle^{(1)}
&=&
\frac{ie^5gB^2}{8\pi^3}\,\lim_{\alpha\to 0^+}
\sum_{\ell=0}^{\infty}\sum_{n=0}^{\infty}e^{-\alpha(\ell+2n)}
\nonumber \\
&\times&
\int dp^0\,\Big\{(\omega+m)^3\,f(\ell+1,n+1;\omega)+(\omega-m)^3\,
f(\ell,n;\omega)
\nonumber \\
&+&
2(n+1)\,eB\,
\big[
  (\omega+m)\,f(\ell+1,n+1;\omega) +
  (\omega-m)\,f(\ell,n+1;\omega)\big]
\Big\}  \; ;
\\
f(\ell,n;\omega)
&\equiv&
(\omega^2-\epsilon_\ell^2)^{-1} (\omega^2-\epsilon_n^2)^{-2}  \; .
\label{f}
\end{eqnarray}

Since the integrand goes to zero at infinity at least as fast as
$(p^0)^{-3}$, one can close the contour depicted in Figure~3 with a
semicircle of infinite radius in the upper half-plane (see Figure~4),
and evaluate the integral using residues. The complete evaluation is
very tedious, but, because of the analytic structure of the integrand,
the result can be written as
\begin{equation}
\label{structure}
\langle j_0\rangle^{(1)}=I_{vac}+I_0\,\theta(-\mu-m)+
\sum_{n=1}^{\infty}\left[I_n\,\theta(\mu-\epsilon_n)+
I_{-n}\,\theta(-\mu-\epsilon_n)\right]  \; .
\end{equation}
Here we shall evaluate $I_0$ explicitly; the other $I'$s can be
evaluated similarly.  $I_0$ results from the contribution of the
pole in $p^0=-\mu-m$ to the integral in Eq.~(\ref{j1a}). The terms
containing such poles are:
\begin{mathletters}
\label{integrals}
\begin{eqnarray}
& &
\int dp^0\, (\omega-m)^3\,f(0,0;\omega) = 0  \; ;
\\
& &
\int dp^0\, (\omega-m)^3\,f(\ell\ne 0,0;\omega) = 2\pi i
\left[
  -\frac{1}{2\ell eB} - \frac{m^2}{\ell^2e^2B^2}
\right]\theta(\mu+m)+\ldots \; ;
\\
& &
\int dp^0\, (\omega-m)^3\,f(0,n\ne 0;\omega) =
\frac{2\pi i\,m^2}{n^2e^2B^2}\,\theta(\mu+m)+\ldots \; ;
\\
& &
\int dp^0\,(\omega-m)\,f(0,n+1;\omega) =
\frac{2\pi i}{4(n+1)^2e^2B^2}\,\theta(\mu+m)+\ldots
\end{eqnarray}
(In order to give a nonvanishing contribution to the integrals
above, $-\mu-m$ must be negative, a condition assured by the factor
$\theta(\mu+m)$. The dots denote the contribution of other poles.)
The total contribution from this pole to $\langle j_0\rangle^{(1)}$
is
\end{mathletters}
\begin{eqnarray}
\label{j1c}
\langle j_0\rangle^{(1)}_{-\mu-m}
&=&
-\frac{e^5gB^2}{4\pi^2}\,\lim_{\alpha\to 0^+}\sum_{n=1}^{\infty}
\left[
  -\frac{e^{-\alpha n}}{2neB}-\frac{m^2e^{-\alpha n}}{n^2e^2B^2}
  +\frac{m^2e^{-2\alpha n}}{n^2e^2B^2}+\frac{e^{-2\alpha(n-1)}}{2neB}
\right]\theta(\mu+m)
\nonumber \\
&=&
-\frac{e^4gB}{8\pi^2}\,\lim_{\alpha\to 0^+}
\left[
  \ln (1-e^{-\alpha})-e^{2\alpha}\,\ln(1-e^{-2\alpha})+\frac{2m^2}{eB}
  \sum_{n=1}^{\infty}\frac{e^{-2\alpha n}-e^{-\alpha n}}{n^2}
\right]\theta(\mu+m)  \; .
\end{eqnarray}
Taking the limit $\alpha\to 0^+$, the first two terms combine to
yield
\begin{equation}
\langle j_0\rangle^{(1)}_{-\mu-m}=
\frac{e^4gB\,\ln 2}{8\pi^2}\,[1-\theta(-\mu-m)]  \; ,
\end{equation}
whereas the third term vanishes (one can take the limit inside the
sum because the latter is uniformly convergent for $\alpha\ge 0$).
It follows that
\begin{equation}
I_0 = -\frac{e^4gB\,\ln 2}{8\pi^2}  \; .
\end{equation}
Performing an analogous calculation for the poles in $p^0=-\mu\pm
\epsilon_n$ $(n=1,2,\ldots)$, one finally finds
\begin{equation}
\label{j_0^1}
\langle j_0\rangle^{(1)}=\frac{e^4gB\,\ln 2}{8\pi^2}\,
\left\{
  \frac{1}{2}-\theta(-\mu-m)+\sum_{n=1}^{\infty}
  \left[
    \theta(\mu-\epsilon_n)-\theta(-\mu-\epsilon_n)
  \right]
\right\}  \; .
\end{equation}
(The condition that the pole in $-\mu-\epsilon_n$ $(n=0,1,2,\ldots)$
must be negative to contribute to the integral in Eq.~(\ref{j1a})
gives rise to a factor $\theta(\mu+\epsilon_n) = 1-
\theta(-\mu-\epsilon_n)$.  Thus, such a pole contributes to $I_{-n}$
{\em and\/} to $I_{vac}$.)

The first order term in the perturbative expansion of
$\langle\sigma_{21}(x)\rangle$ can be obtained with the help of the
following identity~\cite{Zeitlin1,Zeitlin2}, valid with the proviso
that the chemical potential is in an energy gap:
\begin{equation}
\label{sigma-j}
\langle\sigma_{21}\rangle =
  -\frac{\partial}{\partial B}\,\langle j_0\rangle  \; .
\end{equation}
(The reader is invited to check that Eq.~(\ref{sigma-j}) is satisfied
for $V=0$.)  From Eqs.~(\ref{j_0^1}) and (\ref{sigma-j}) one
immediately obtains
\begin{equation}
\langle\sigma_{21}\rangle^{(1)}=
-\frac{e^4g\,\ln 2}{8\pi^2}
\left\{
  \frac{1}{2}-\theta(-\mu-m)+\sum_{n=1}^{\infty}
  \left[
    \theta(\mu-\epsilon_n)-\theta(-\mu-\epsilon_n)
  \right]
\right\}  \; .
\end{equation}
This result agrees with the (much harder) direct computation of
$\langle\sigma_{21}\rangle^{(1)}$.

\section{Correlated disorder}
%        -------------------
\label{FiniteCorrelation}

Now, let us repeat the calculation of the previous section with a
modified probability distribution for $V$:
\begin{equation}
\label{P'}
P[V]=\exp
\left\{
  -\frac{1}{2gM^2}\int d^2x\,
  \left[(\nabla V)^2 + M^2V^2\right]
\right\}  \; ,
\end{equation}
for which
\begin{equation}
\langle V({\bf x})V({\bf y})\rangle
=
gM^2\int\frac{d^2p}{(2\pi)^2}\,
\frac{e^{i{\bf p}\cdot({\bf x}-{\bf y})}}{p^2+M^2}
=
\frac{gM^2}{4\pi}\int_0^{\infty}\frac{ds}{s}\,
e^{-M^2s-({\bf x}-{\bf y})^2/4s}  \; .
\end{equation}
Substituting the expression above into the first line of
Eq.~(\ref{j1}), integrating over $y^0$, $z^0$, $q^0$ and $k^0$,
and defining new integration variables ${\bf u}={\bf x}-{\bf y}$
and ${\bf v}={\bf z}-{\bf x}$, one obtains
\begin{eqnarray}
\label{j1e}
\langle j_0\rangle^{(1)}
&=&
i\pi e^3gM^2 \int_0^{\infty} \frac{ds}{s}\,
e^{-M^2s} \int d^2u\,d^2v \int dp^0\,\exp
\left\{
  -\frac{ieB}{2}\,\epsilon_{ij}u^iv^j
  -\frac{({\bf u}+{\bf v})^2}{4s}
\right\}
\nonumber \\
&\times&
{\rm Tr}
\left[
  \gamma_0\,\Sigma(p^0,{\bf u})\,\gamma_0\,
  \Sigma(p^0,-({\bf u}+{\bf v}))\,\gamma_0\,\Sigma(p^0,{\bf v})
\right]  \; .
\end{eqnarray}
Evaluating the trace and performing the integrals over {\bf u} and
{\bf v} (see Appendix \ref{app_int}), one finds ($z=2eBs$)
\begin{eqnarray}
\label{j1b}
\langle j_0\rangle^{(1)}
&=&
\frac{ie^4gBM^2}{16\pi^3}\,\lim_{\alpha\to 0^+}\,
\int_0^{\infty} dz\,e^{-M^2z/2eB}\,\sum_{\ell=0}^{\infty}
\sum_{n=0}^{\infty} e^{-\alpha(\ell+2n)}
\nonumber \\
&\times&
\int dp^0\,
\Big\{
  \phi(\ell,n,n;z)
  \left[
    (\omega+m)^3 f(\ell+1,n+1;\omega)+
    (\omega-m)^3 f(\ell,n;\omega)
  \right]
\nonumber \\
&+&
  4(n+1)\,eBz\,\phi(\ell,n,n+1;z)
  \left[
    (\omega+m)+e^{-\alpha}\,(\omega-m)
  \right]
  f(\ell+1,n+1;\omega)
\nonumber \\
&+&
  2(n+1)\,eB
  \left[
    \phi(\ell,n,n;z)\,(\omega+m)\,f(\ell+1,n+1;\omega)+
    \phi(\ell,n+1,n+1;z)\,(\omega-m)\,f(\ell,n+1;\omega)
  \right]
\Big\}  \; ,
\end{eqnarray}
where
\begin{equation}
\phi(\ell,m,n;z) \equiv \sum_{k=0}^{m}\pmatrix{m\cr k\cr}
\pmatrix{\ell-k+n\cr n\cr}\,\frac{(z-1)^{k}}{(z+1)^{\ell-k+n+1}}  \; ,
\end{equation}
and $f(\ell,n;\omega)$ is defined in Eq.~(\ref{f}).

Here, too, the analytic structure of the integrand in Eq.~(\ref{j1b})
implies that $\langle j_0\rangle^{(1)}$ has the form shown in
Eq.~(\ref{structure}).  As in the previous section, we shall evaluate
only $I_0$.  The terms containing a pole in $p^0=-\mu-m$ are the same
as in Eq.~(\ref{integrals}); substituting those results into
Eq.~(\ref{j1b}), one obtains
\begin{eqnarray}
\label{j1d}
\langle j_0\rangle^{(1)}_{-\mu-m}
&=&
-\frac{e^4gBM^2}{8\pi^2}\,
\lim_{\alpha\to 0^+} \int_0^{\infty} dz\,\sum_{n=1}^{\infty}
\frac{e^{-M^2z/2eB}}{(z+1)^{n+1}}
\left[
  -\frac{e^{-\alpha n}}{2neB}-\frac{m^2e^{-\alpha n}}{n^2e^2B^2}
  +\frac{m^2e^{-2\alpha n}}{n^2e^2B^2}+\frac{e^{-2\alpha(n-1)}}{2neB}
\right]\theta(\mu+m)
\nonumber \\
&=&
-\frac{e^3gM^2}{16\pi^2}\,
\lim_{\alpha\to 0^+}\int_0^{\infty} dz\,\frac{e^{-M^2z/2eB}}{z+1}\,
\Bigg[
  \ln\left( 1-\frac{e^{-\alpha}}{z+1} \right)
  -e^{2\alpha}\,\ln\left( 1-\frac{e^{-2\alpha}}{z+1} \right)
\nonumber \\
&+&
  \frac{2m^2}{eB}\,\sum_{n=1}^{\infty}
  \frac{e^{-2\alpha n}-e^{-\alpha n}}{n^2(z+1)^n}\,
\Bigg]\theta(\mu+m)  \; .
\end{eqnarray}
The expression above (and, therefore, $I_0$) {\em vanishes\/} for
any finite $M$ in the limit $\alpha\to 0^+$ (see
Appendix \ref{app_proof} for a proof).  On the other hand, if the
limit $M\to\infty$ is taken first, then the result of the previous
section is recovered: in fact, if $f(z)$ is a function continuous at
the origin, then
\begin{equation}
\lim_{M\to\infty}\,M^2 \int_0^{\infty} dz\,e^{-M^2z/2eB}\,f(z)
=
2eB \lim_{M\to\infty}\, \int_0^{\infty} dx\,e^{-x}\,f(2eBx/M^2)
=
2eB\,f(0)  \; .
\end{equation}
Using this result in Eq.~(\ref{j1d}), one recovers Eq.~(\ref{j1c}).

\section{Conclusions}
%        -----------
\label{Conclucoes}

The results of the two preceding sections represent a first attempt
at a systematic study of the effects of disorder on the 2DEG in a
uniform magnetic field.

The perturbative treatment adopted, suitable for weak disorder,
is incapable of generating the plateaux which appear in the
(integer-) Quantized Hall Effect, when one plots transverse
conductivity versus charge density.  In fact, for both correlated
and uncorrelated disorder, one still obtains a sequence of points
lying along a straight line that satisfies the classical relation
of Eq.~(\ref{classical}).

A comparison between the cases of uncorrelated and correlated
disorder shows that, in the former, both charge density and
transverse conductivity are altered by perturbative corrections,
whereas in the latter (more physical) situation, this does not
seem to occur (at least in a first order calculation). Based
on our explicit calculation of $I_0$ and $I_{\pm 1}$ (not
shown here), we
conjecture that this feature will hold for all Landau levels
(i.e., $I_n=0$ for all $n$).
This would indicate that the values of transverse conductivity
and charge density would be indistinguishable from those in
the absence of disorder.

It is important to note that our perturbative results show that
the Landau levels remain degenerate, the weak disorder being unable
to break this degeneracy and broaden them into bands.  Also, no
localized states appear in the spectrum.  As these are often claimed
to be essential ingredients for plateaux formation, it should not be
surprising that our results reproduce the classical behavior.

In closing, we would like to mention that such features may be
present if one sums up the contribution of an infinite subclass
of graphs in the perturbative expansion.  This will induce
non-perturbative corrections to the electron propagator and lead
to a broadening of the Landau levels into bands and to the
appearance of interband states.  We have obtained preliminary
results in this direction, which make us hope that this might lead
to qualitative changes in the graph of transverse conductivity
versus charge density.  We hope to report on that work shortly.

\acknowledgments
%---------------
%
This work had financial support from Proyecto FONDECYT, under Grant
No.\ 1950794, Fundaci\'on Andes, FUJB/UFRJ, CNPq, FAPERJ, UNESCO and
CLAF.  R.M.C.\ and C.A.A.C.\ thank PUC-Chile, N.B. thanks UFRJ, and
P.D.\ thanks UFRJ and CLAF, all for the kind hospitality received
while this work was in progress.

\appendix

\section{}
%--------
\label{app_int}

The integrals over {\bf u} and {\bf v} in Eq.~(\ref{j1e}) have the
form
\begin{equation}
\label{intuv}
\int d^2u\,d^2v\,e^{-F({\bf u},{\bf v})}\,
P(u^i,v^j)\,L_{\ell}^{\alpha}\big(\frac{eB}{2}\,{\bf u}^2\big)\,
L_{m}^{\beta}\big(\frac{eB}{2}\,({\bf u}+{\bf v})^2\big)\,
L_{n}^{\gamma}\big(\frac{eB}{2}\,{\bf v}^2\big)  \; ,
\end{equation}
where
\begin{equation}
F({\bf u},{\bf v}) = \frac{ieB}{2}\,\epsilon_{ij}u^iv^j+
\frac{1}{4s}\,({\bf u}+{\bf v})^2+
\frac{eB}{2}\,({\bf u}^2+{\bf u}\cdot{\bf v}+{\bf v}^2)  \; ,
\end{equation}
$P(u^i,v^j)$ is some polynomial in $u^i$, $v^j$, and
$L_{\ell}^{\alpha}(x)$ is a Laguerre polynomial.  Here we show
how to evaluate such integrals.

With the help of the generating function of the Laguerre
polynomials\cite{GR},
\begin{equation}
(1-t)^{-1-\alpha}\,\exp\left(\frac{tx}{t-1}\right)=
\sum_{n=0}^{\infty}L_n^{\alpha}(x)\,t^n\qquad(|t|<1)  \; ,
\end{equation}
we define another generating function,
\begin{eqnarray}
{\cal Z}(x,y,z;{\bf p},{\bf q})
&\equiv&
\sum_{\ell=0}^{\infty} \sum_{m=0}^{\infty} \sum_{n=0}^{\infty}
\int d^2u\,d^2v\,
e^{{\bf p}\cdot{\bf u}+{\bf q}\cdot{\bf v}-F({\bf u},{\bf v})}\,
L_{\ell}^{\alpha} \big(\frac{eB}{2}\,{\bf u}^2\big)\,
L_{m}^{\beta} \big(\frac{eB}{2}\,({\bf u}+{\bf v})^2\big)\,
L_{n}^{\gamma} \big(\frac{eB}{2}\,{\bf v}^2\big)\, x^{\ell}y^mz^n
\nonumber \\
&=&
\frac{4\pi^2 S^{-1}}{(1-x)^{\alpha}(1-y)^{\beta}(1-z)^{\gamma}}\,
\exp
\left\{
  \frac{(1-x)(1-y)(1-z)}{S}
  \left[
    a\,{\bf p}^2 +b\,{\bf q}^2+c\,{\bf p}\cdot{\bf q}+
    d\,\epsilon_{ij}p^iq^j
  \right]
\right\}  \; ,
\end{eqnarray}
where
\begin{mathletters}
\begin{eqnarray}
a&=&\frac{1}{4s}+\frac{eB}{2}\frac{(1-yz)}{(1-y)(1-z)}  \; , \\
b&=&\frac{1}{4s}+\frac{eB}{2}\frac{(1-xy)}{(1-x)(1-y)}  \; , \\
c&=&-\frac{1}{2s}-\frac{eB}{2}\left(\frac{1+y}{1-y}\right)  \; , \\
d&=&-\frac{ieB}{2}  \; , \\
S&=&e^2B^2(1-xyz)+\frac{eB}{2s}\,(1-y)(1-xz)\qquad(0\le x,y,z<1)\; .
\end{eqnarray}
\end{mathletters}
Now, the integrals in Eq.~(\ref{intuv}) can be obtained by expanding
the following expression in a power series in $x$, $y$ and $z$:
\begin{equation}
P\big(\frac{\partial}{\partial p^i},\frac{\partial}{\partial q^j}\big)\,
{\cal Z}(x,y,z;{\bf p},{\bf q})\big|_{{\bf p}={\bf q}={\bf 0}}  \; .
\end{equation}

\section{}
%--------
\label{app_proof}

Here we show that $\langle j_0\rangle^{(1)}_{-\mu-m}=0$, for any
finite $M$.  To prove this, it is convenient separate the integral
in Eq.~(\ref{j1d}) in three pieces. The first one we consider is
\begin{eqnarray}
I_1
&\equiv&
-\frac{e^2gM^2m^2}{8\pi^2B}\,\lim_{\alpha\to 0^+}\,
\int_0^{\infty} dz\,e^{-M^2z/2eB}\,\sum_{n=1}^{\infty}
\frac{e^{-2\alpha n}-e^{-\alpha n}}{n^2(z+1)^{n+1}}
\nonumber \\
&\le&
\frac{e^2gM^2m^2}{8\pi^2B}\,\lim_{\alpha\to 0^+}\,
\int_0^{\infty} dz\,\sum_{n=1}^{\infty}
\frac{e^{-\alpha n}-e^{-2\alpha n}}{n^2(z+1)^{n+1}}
\nonumber \\
&=&
\frac{e^2gM^2m^2}{8\pi^2B}\,\lim_{\alpha\to 0^+}\,
\sum_{n=1}^{\infty}\frac{e^{-\alpha n}-e^{-2\alpha n}}{n^3}=0  \; .
\end{eqnarray}
Since $I_1\ge 0$, it follows that $I_1=0$.  The next piece we
consider is
\begin{eqnarray}
I_2
&\equiv&
-\frac{e^3gM^2}{16\pi^2}\,\lim_{\alpha\to 0^+}
\int_0^{\infty} dz\,\frac{e^{-M^2z/2eB}}{z+1}\,
\left[
  \ln \left(1-\frac{e^{-\alpha}}{z+1}\right) -
  \ln \left(1-\frac{e^{-2\alpha}}{z+1}\right)
\right]
\nonumber \\
&=&
\frac{e^3gM^2}{16\pi^2}\,\lim_{\alpha\to 0^+}
\left(\int_0^c+\int_c^{\infty}\right) dz\,\frac{e^{-M^2z/2eB}}{z+1}\,
\ln \left(\frac{z+1-e^{-2\alpha}}{z+1-e^{-\alpha}}\right)
\equiv
I_2^1+I_2^2  \; ;
\\
I_2^1
&\le&
\frac{e^3gM^2}{16\pi^2}\,\lim_{\alpha\to 0^+}\,c\,
\ln \left(\frac{1-e^{-2\alpha}}{1-e^{-\alpha}}\right)
=
\frac{e^3gM^2}{16\pi^2}\,c\,\ln 2  \; ;
\\
I_2^2
&\le&
\frac{e^3gM^2}{16\pi^2}\,\lim_{\alpha\to 0^+}\,\frac{1}{c+1}\,
\ln \left(\frac{c+1-e^{-2\alpha}}{c+1-e^{-\alpha}}\right)
\int_c^{\infty} dz\,e^{-M^2z/2eB}
\nonumber \\
&=&
\frac{e^4gB}{8\pi^2}\,\lim_{\alpha\to 0^+}\,
\frac{e^{-M^2c/2eB}}{c+1}\,
\ln \left(\frac{c+1-e^{-2\alpha}}{c+1-e^{-\alpha}}\right)
= 0  \; .
\end{eqnarray}
Since $c$ can be chosen arbitrarily small, it follows that $I_2=0$.
The remaining piece is
\begin{eqnarray}
I_3
&\equiv&
-\frac{e^3gM^2}{16\pi^2}\,\lim_{\alpha\to 0^+}
\int_0^{\infty} dz\,e^{-M^2z/2eB}\,\frac{e^{2\alpha}-1}{z+1}\,
\ln \left(1-\frac{e^{-2\alpha}}{z+1}\right)
\nonumber \\
&\le&
-\frac{e^4gB}{8\pi^2}\,\lim_{\alpha\to 0^+}\,
(e^{2\alpha}-1)\,\ln\,(1-e^{-2\alpha})
= 0  \; .
\end{eqnarray}
This completes the proof.

% FIGURES
% -------
\newpage

\noindent
\underline{\bf Figure Captions}:

\vspace{5mm}
\noindent
{\bf Figure 1}:
Transverse conductivity $\sigma$ (in units of $e^2/h$) vs. electron
density $\rho$ (in units of $eB/h$) in the absence of disorder.  The
dashed line is the classical relation, while the dots correspond to
fully occupied Landau levels, the only values attainable by varying
the chemical potential when the levels have zero width.  The shift of
both axes by half a unit is characteristic of the relativistic theory.

\vspace{5mm}
\noindent
{\bf Figure 2}: (a) Feynman diagrams for $\langle j_0(x)\rangle$
up to first order in $g$; (b) the same for $\langle\Pi_{\mu\nu}(x,y)\rangle$.
A fermion propagator is associated with each solid line; a two-point
correlation function of the random potentials with each dashed line;
a factor $e\gamma_0$ with each vertex connecting two solid and one dashed
lines; a factor $e\gamma_{\mu}$ with each cross.

\vspace{5mm}
\noindent
{\bf Figure 3}: Integration contour in the complex $p^0$-plane used
in the definition of the Feynman propagator.

\vspace{5mm}
\noindent
{\bf Figure 4}: Integration contour in the complex $p^0$-plane used
in the computation of the integral in Eq.\ (\ref{j1a}).


\begin{references}
%      ----------

\bibitem[a)]{NB} E-mail: nbralic@lascar.puc.cl

\bibitem[b)]{RMC} E-mail: rmc@fis.puc-rio.br

\bibitem[c)]{CAAC} E-mail: aragao@if.ufrj.br

\bibitem[d)]{PD} E-mail: sabrina@shannon.sissa.it

\bibitem{QED3} J. Schwinger, Phys. Rev. {\bf 82}, 664 (1951);
  A. N. Redlich, Phys. Rev. D {\bf 29}, 2366 (1984);
  V. P. Gusynin, V. A. Miransky and I. A. Shovkovy,
    Phys. Rev. Lett. {\bf 73}, 3499 (1994); Phys. Rev. D {\bf 52}, 4718
    (1995);
  Y. Hosotani, Phys. Lett. {\bf B319}, 332 (1993); Phys. Rev. D
    {\bf 51}, 2022 (1995).

\bibitem{Chodos} A. Chodos, K. Everding, D. A. Owen, Phys. Rev. D {\bf 42},
    2881 (1990);
  P. Elmfors, D. Persson, B-S. Skagerstam, Phys. Rev. Lett.
    {\bf 71}, 480 (1993).

\bibitem{IQHE} For reviews see, for example,
  K. von Klitzing, Rev. Mod. Phys. {\bf 58}, 519 (1986);
  M. Jan{\ss}en, O. Viehweger, U. Fastenrath and J. Hajdu,
    {\it Introduction to the Theory of the Integer Quantum Hall
        Effect} (VCH Verlagsgesellschaft, Weinheim, 1994).

\bibitem{Lykken} J. D. Lykken, J. Sonnenschein and N. Weiss,
  Int. J. Mod. Phys. {\bf A6}, 1335 and 5155 (1991).

\bibitem{Zeitlin1} V. Zeitlin, Phys. Lett. {\bf B352}, 422 (1995).

\bibitem{Zeitlin2} V. Zeitlin, Mod. Phys. Lett. {\bf A12}, 877 (1997).

\bibitem{DKK} D. K. Kim and K.-S. Soh, Phys. Rev. D {\bf 55}, 
  6218 (1997).

\bibitem{GR} I. S. Gradshteyn and I. M. Ryzhik, {\it Table of
  Integrals, Series, and Products\/}, 4th ed. (Academic Press, New
  York, 1965).

\bibitem{IZ} C. Itzykson and J.-B. Zuber, {\it Quantum Field
  Theory\/} (McGraw-Hill, Singapore, 1980).

%\bibitem{weak} We were unable to prove this. However, we have checked
%  that $I_{\pm 1}=0$.

%\bibitem{Tarrach} R. Gosdzinsky and R. Tarrach, Am. J. Phys. {\bf 59}, 70
%  (1991).

\end{references}
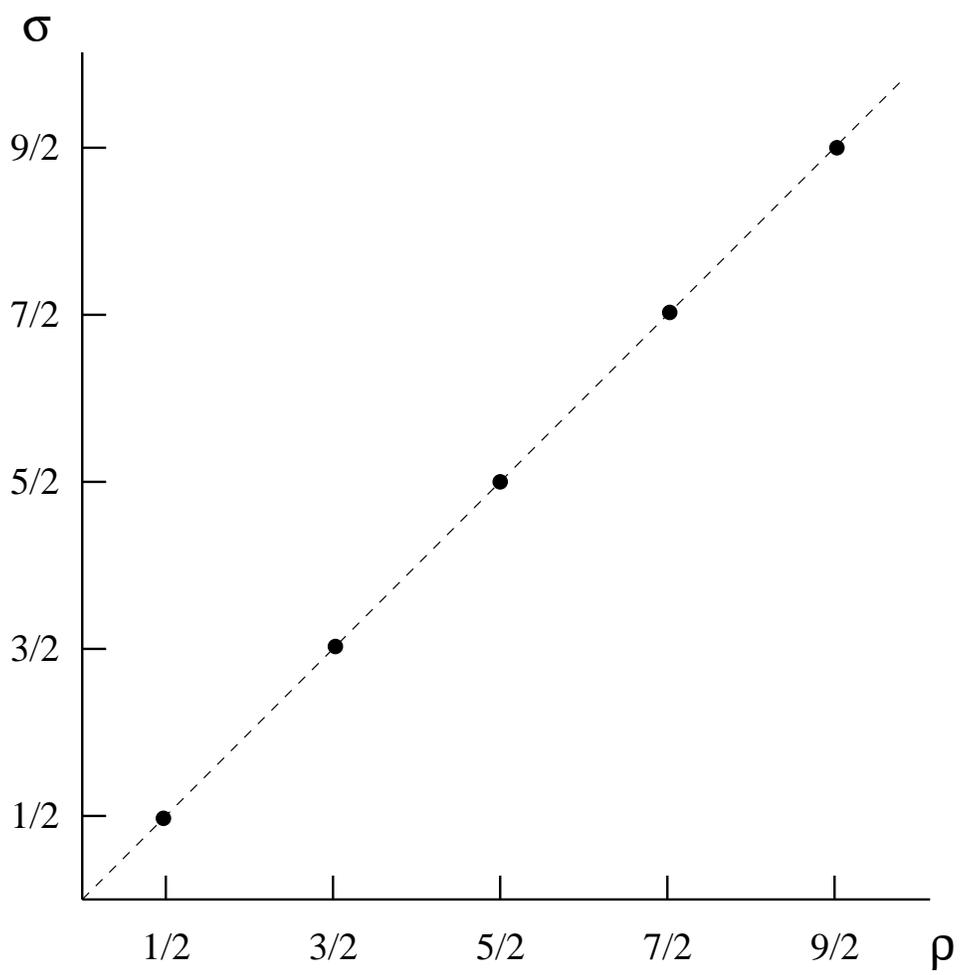
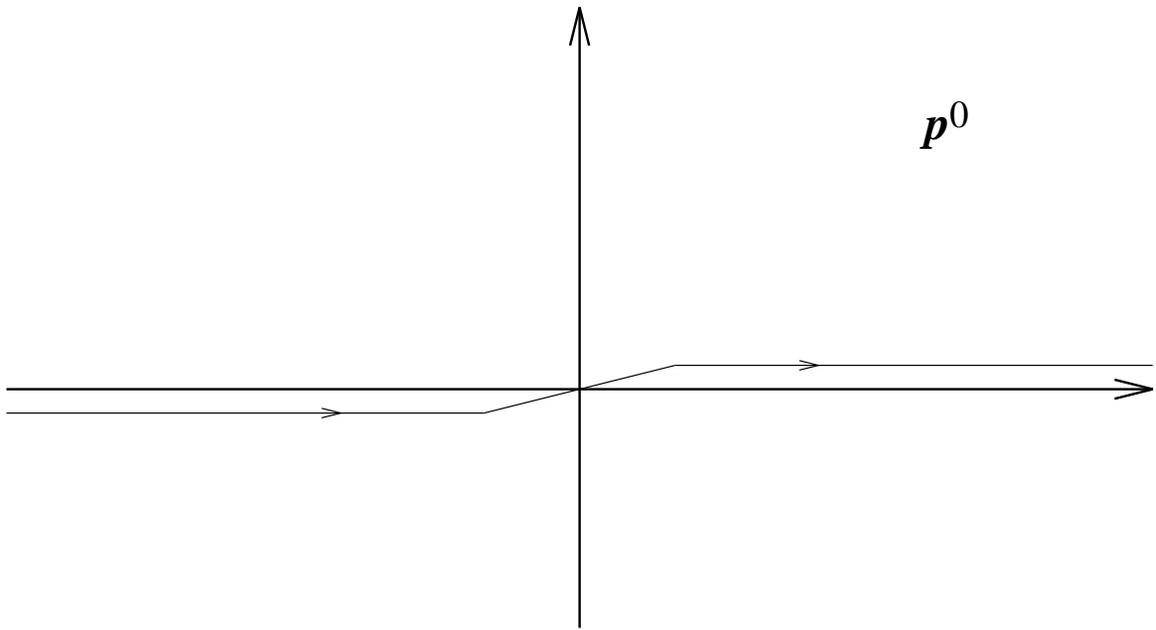
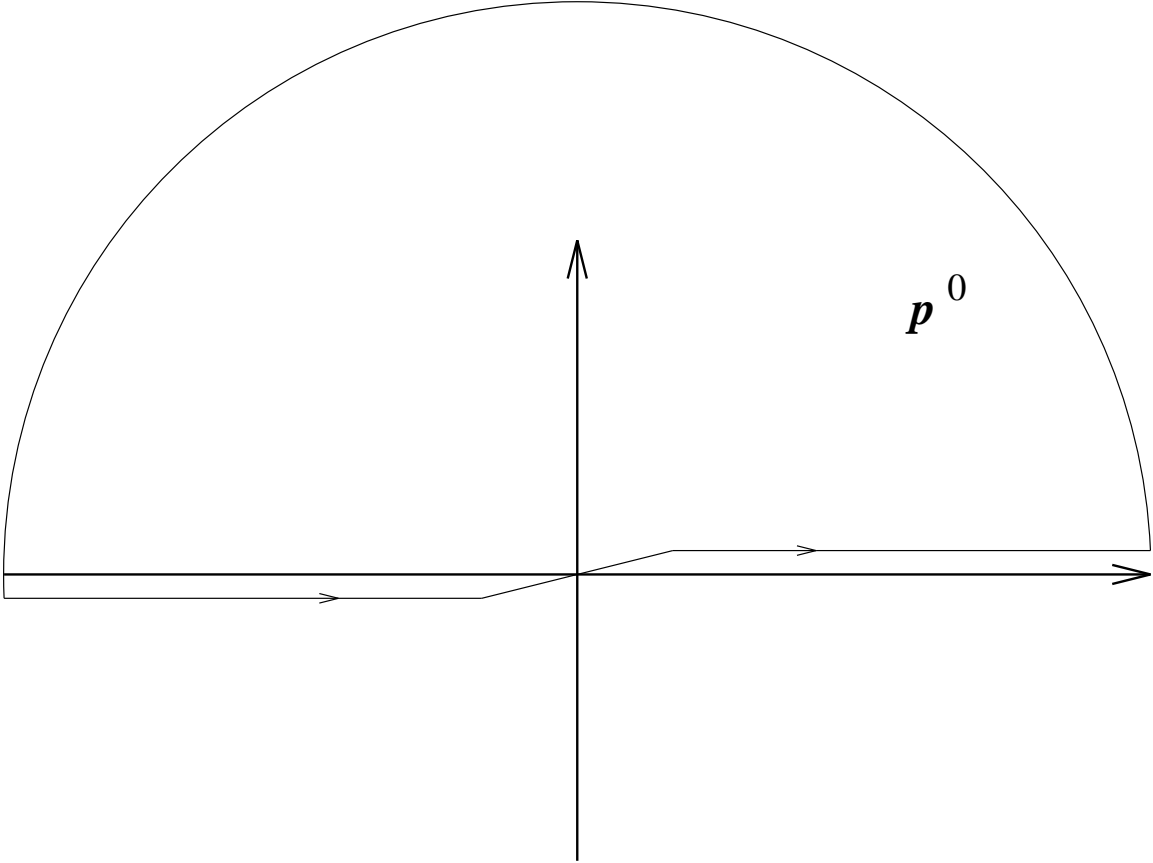
\end{document}